\documentclass[12pt]{iopart}

\usepackage{iopams}
\usepackage{graphicx}
\usepackage[toc,page,title,titletoc,header]{appendix}
\begin{document}
\bibliographystyle{unsrt}

\title{Higgs algebraic symmetry of screened system in a spherical geometry}
\author{Yan Li$^{1,3}$, Fu-Lin Zhang$^{2}$$^{\dag}$, and Jing-Ling Chen$^{1,3}$$^{\ddag}$}

\address{$^{1}$Theoretical Physics Division, Chern Institute of
Mathematics, Nankai University, Tianjin 300071, People's Republic of
China \\ $^{2}$Physics Department, School of Science, Tianjin
University, Tianjin 300072, People's Republic of China
\\ $^{3}$Centre for Quantum Technologies, National University of Singapore, 3 Science Drive 2, Singapore 117543} \ead{{\dag}
flzhang@tju.edu.cn, {\ddag} chenjl@nankai.edu.cn }

\begin{abstract}
The orbits and the dynamical symmetries for the screened Coulomb potentials and isotropic harmonic oscillators have been studied by Wu and Zeng [Z. B. Wu and J. Y. Zeng, Phys. Rev. A 62,032509 (2000)]. We find the similar properties in the responding systems in a spherical space, whose dynamical symmetries are described by Higgs Algebra. There exists a conserved
aphelion and perihelion vector, which, together with angular momentum, constitute the generators of the geometrical symmetry group
at the aphelia and perihelia points $(\dot{r}=0)$.
\end{abstract}



\maketitle


\section {Introduction}\label{intro}

The classic Bertrand's theorem says that
there are two central fields only for which all bounded orbits are closed, namely,
the Kepler's
problem (also known as the inverse square law) and the
 isotropic harmonic oscillator
\cite{Bertrand,Goldstein,santos2011english}.
The both fields give rise to elliptical orbits,
with the difference that in the first case the force is directed towards one of the foci and in the
second case the force is directed
to the geometrical center of the ellipse.
Besides the energy and angular momentum,
these elliptical orbits are guaranteed by an additional conserved
quantity: the Runge-Lenz vector
for Kepler's
problem \cite{Runge,landau1975classical,Lenz,ngome2009curved} and a quadrupole tensor
for the isotropic harmonic oscillator \cite{Jauch,elliott1958collective}, which imply higher
dynamical symmetries than the geometric symmetries \cite{Pauli,Schiff,Zeng}.
In this paper, we pay attention to the two-dimensional cases, in which
the angular momentum $\textbf{L}=L_z\textbf{k}$,
the Runge-Lenz vector has two components
\begin{eqnarray}\label{R1}
R_x&=&\frac{1}{2}(p_yL_z+L_zp_y)-\frac{x}{r},\\
R_y&=&-\frac{1}{2}(p_xL_z+L_zp_x)-\frac{y}{r},\nonumber
\end{eqnarray}and
the components of the conserved tensor in oscillator are
\begin{eqnarray}
Q_{xy}=xy+p_xp_y,\\
Q_1=\frac{1}{2}[(x^2-y^2)+(p_x^2-p_y^2)].\nonumber
\end{eqnarray}

Taking the classical orbit as his start point, Higgs \cite{higgs1979dynamical} generalized
the hydrogen atom and harmonic oscillator in the spherical space preserving the dynamical symmetry. He established a
gnomonic projective coordinate system in which the orbit of the motion on a sphere can be described
as
\begin{eqnarray}\label{guiji}
\frac{1}{2}L_z^2[r^{-4}(\frac{dr}{d\theta})^2+r^{-2}]+V(r)=E-\frac{1}{2}\lambda
L_z^2,
\end{eqnarray}
where the angular momentum $L_z=x_1p_2-x_2p_1$ is an invariant
quantity with the radial symmetric potential $V(r)$. The curvature appears only in the right combination
$E-\frac{1}{2}\lambda L_z^2$ of Eq. (\ref{guiji}), and, therefore, the projected
orbits are the same, for a given $V(r)$, as in Euclidean geometry.

To give a more clear description of Higgs' result, we shall review the coordinate systems adopt in \cite{higgs1979dynamical}.
\begin{figure}
\includegraphics[width=10cm]{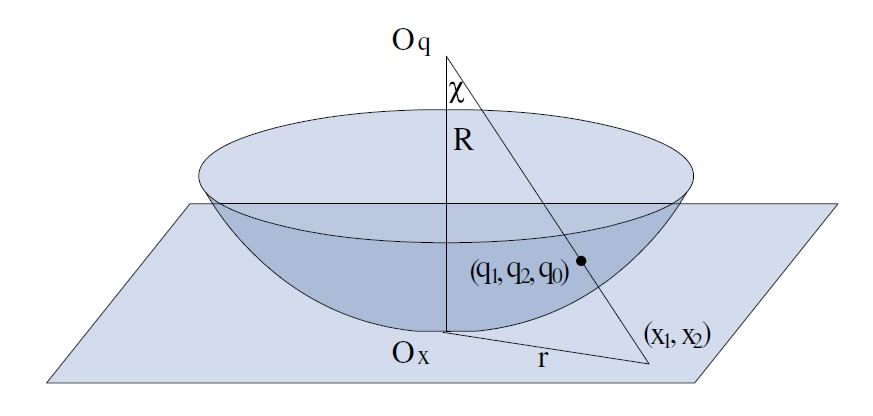}\caption{The gnomonic projection,
which is the projection onto the tangent plane from the center of
the sphere in the embedding space, and the coordinate systems}\label{fig}
\end{figure}
Consider the sphere embed in a three-dimensional  Euclidean space, see Fig. \ref{fig}.
Then, each
pair of independent variables $(q_1, q_2)$ of the three-dimensional
Cartesian coordinates $(q_1, q_2,q_0)$, with the origin $O_q$ in the
figure, is obtained simply by imposing the constraint
\begin{eqnarray}
q_0^1+q_1^2+q_2^2=\frac{1}{\lambda},
\end{eqnarray}
where $\lambda=\frac{1}{R^2}$ is the
curvature of the sphere, in which R is the radius.
The points on the sphere can also be
described by the spherical coordinates $(R, \chi,\theta)$
which is defined by $(q_1, q_2,q_0)=(R\sin\chi\cos\theta,
R\sin\chi\sin\theta, R\cos\chi)$. On the other hand, the points can be expressed in terms of polar coordinates $(r,\theta)$.
In the two-dimensional gnomonic projection, which is the projection onto the tangent plane from the center of the sphere in the embedding space, the Cartesian coordinates of this projection, denoted $(x_1, x_2)$, are given by
\begin{eqnarray}
q_1=\frac{x_1}{\sqrt{1+\lambda r^2}},\ \ \ \ \ \ \ \ \
q_2=\frac{x_2}{\sqrt{1+\lambda r^2}},
\end{eqnarray}
where $r^2=x_1^2+x_2^2$ and the point of tangency $O_x$ in the
figure is the origin. Then, the polar coordinates $(r,\theta)$ of
the projection are determined by the relations  $r=R\tan\chi$ and $(x_1,
x_2)=(r\cos\theta, r\sin\theta)$.


Higgs demonstrated that in the gnomonic projection, the Hamiltonian in a spherical geometry, with the same classical orbits corresponding in a plane,    can be written as
\begin{eqnarray}\label{Hamilton}
H=\frac{\pi^2}{2}+\frac{1}{2}\lambda L_z^2+V(r),
\end{eqnarray}
where
$\vec{\pi}=\vec{p}+\frac{\lambda}{2}\left[\vec{x}(\vec{x}\cdot\vec{p})+(\vec{p}\cdot\vec{x})\vec{x}\right]$
is the conserved vector in free particle motion on the sphere.
 He gave the conserved quantities which commute with the Hamiltonian for the Coulomb potential and the isotropic oscillator.
According to the Bertrand's Theorem, the orbits are closed only if the
potential takes the Coulomb or isotropic oscillator form, i.e.
$V(r)=-\frac{1}{r}$ or $V(r)=\frac{1}{2}r^2$. The systems
described by Eq. (\ref{Hamilton}) with the two mentioned potentials
are defined as the Kepler problem and isotropic oscillator in a
spherical geometry in \cite{higgs1979dynamical}. The algebraic
relations of their conserved quantities reveal the dynamical
symmetries of the two systems are described by the symmetry groups $SO(3)$ and
$SU(2)$ respectively.  This algebraic structure is called \emph{Higgs Algebra}, and
 has received attention from a
variety of literatures
\cite{bacry1966dynamical,KK,zhang2009higgs,JLChenHiggs,CS,li2011virial,esteve2012boundary}.

The Bertrand's theorem has been extended in different directions
\cite{wu2000dynamical,zeng2002closure,wu1998extension,zuo1999modification}. In the original
theorem, the form of the central potential is assumed to be a power-law function of $r$ \cite{Goldstein}.
However,  Wu and Zeng \cite{wu2000dynamical} believed that, if the restriction of a power-law form of
the central potential is relaxed, Bertrand's theorem may be extended. They showed
that when the Coulomb potential or
isotropic harmonic oscillator is screened [see Eq. (\ref{coulomb}) and Eq. (\ref{harmonic})], the elliptic orbits are broken, but there still exist an infinite
number of closed orbits. The broken orbit gives promise of that, the dynamical symmetry
$SO(3)$ for hydrogen atom or $SU(2)$ for isotropic harmonic oscillator is broken, but the revival of closeness of some classical
orbits may be an indication of the recurrence of the dynamical
symmetry.

For the screened two-dimensional Coulomb potential,  Wu and Zeng \cite{wu2000dynamical} noticed that the usual
Runge-Lenz vector in Eq. (\ref{R1}) no longer remains conserved. While, they found that
 the extended
Runge-Lenz vector with components given bellow
\begin{eqnarray}\label{R2}
R'_x&=&\frac{1}{2}(p_yL_z+L_zp_y)-\left(1+\frac{2k}{r}\right)\frac{x}{r},\\
R'_y&=&-\frac{1}{2}(p_xL_z+L_zp_x)-\left(1+\frac{2k}{r}\right)\frac{y}{r},\nonumber
\end{eqnarray}
are conserved at the aphelion (perihelion) points ($\dot{r}=0$).
In other words, besides $\left[L_z,H\right]=0$,  it holds that $\left[R'_x,H\right]=0$ and$ \left[R'_y,H\right]=0$ at the aphelion and perihelion points.
Moreover, they satisfy the commutation relations
\begin{eqnarray}
\left[L_z,R'_x\right]&=&iR'_y,\\
\left[L_z,R'_y\right]&=&-iR'_x,\nonumber\\
\left[R'_x,R'_y\right]&=&-2iHL_z,\nonumber
\end{eqnarray}
which imply that ($L_z, R'_x, R'_y$) constitute an $SO(3)$ algebra in
Hilbert space spanned by degenerate states belonging to a
given energy eigenvalue $E_n=-\frac{1}{2n^2}$. Here,
$n=n_r+|m'|+\frac{1}{2}$, where
 $n_r=0,1,2,\ldots$ and $m'=\sqrt{m^2-2k}$ with $m=-n_r,-n_r+1,\ldots,n_r$ is the eigenvalue of the conserved angular
momentum $L_z$. The authors of \cite{wu2000dynamical} believed, in
general, that the dynamical symmetry $SO(3)$ of a two-dimensional hydrogen
atom is broken, and the $SO(3)$ symmetry may be restored at
the aphelion and perihelion points of the classical orbits. Similar results have been obtained in the screened two-dimensional isotropic oscillator.

Since the construction of the systems in the sphere given by Higgs \cite{higgs1979dynamical} is based on the orbits of classical motion, which is also the starting point in the work
 of Wu and Zeng \cite{wu2000dynamical}, we speculate that  similar results of the screened potentials in \cite{wu2000dynamical} can be found in Higgs' sphere systems  \cite{higgs1979dynamical}.
%
%
 As the first attempt, we focus on two-dimensional
spherical geometry. In Sec. \ref{coulomb potential}, we will show the dynamical symmetry of screened Coulomb potential
to a spherical geometry. Besides that, we give the eigenenergy and eigenstates of this system. A similar process for the screened isotropic harmonic oscillator will be presented in Sec.
\ref{isotropic harmonic}. A summary is made in the last section.

\section{Screened Coulomb potential}\label{coulomb potential}


Based on the results in \cite{wu2000dynamical}, the classical orbit equation in the spherical system described by the Hamiltonian (\ref{Hamilton}) with the screened Coulomb potential
\begin{eqnarray}\label{coulomb}
V(r)=-\frac{1}{r}-\frac{k}{r^2}\ \ \ \ \ \ (0<k\ll1),
\end{eqnarray}
can be obtained by using the relation in Eq. (\ref{guiji}) as
\begin{eqnarray}\label{re-1}
\frac{1}{r}=\frac{1}{L_z^2 \alpha^2}\left[1+\sqrt{1+2(E-\frac{1}{2}\lambda L_z^2)L_{z}^2 \alpha^2}\cos{(\alpha(\theta-\theta_0))}\right],
\end{eqnarray}
where $E<0$ and $\alpha=\sqrt{1-2k/L_z^2}<1$. When $\lambda\rightarrow0$, the equation
reduces to the screened Coulomb potential on a plane.

%
   It should be noted that the  orbit is not closed in general.
However, for rational values of $\alpha$  (that is,  for suitable angular
momenta $L_z=\sqrt{2k/(1-\alpha^2)}$), there still exist an infinite
number of closed orbits  whose geometry depends only on the angular momentum.


In the construction of Higgs, he replaced $\vec{p}$ in the original Runge-Lenz vector by $\vec{\pi}$, which is conserved in free motion in a spherical geometry, and obtained the Runge-Lenz vector in this space
\begin{eqnarray}
R'_x=\frac{1}{2}\left(\pi_yL_z+L_z\pi_y\right)-\left(1+\frac{2k}{r}\right)\frac{x}{r},\\
R'_y=-\frac{1}{2}\left(\pi_xL_z+L_z\pi_x\right)-\left(1+\frac{2k}{r}\right)\frac{y}{r}.\nonumber
\end{eqnarray}
By a direct calculation, we find that they are still conserved at the aphelion (perihelion) points ($\dot{r}=0$), i.e.,
\begin{eqnarray}\label{rh}
\left[R'_x,H\right]=0,\ \ \ \ \ \ \left[R'_y,H\right]=0.
\end{eqnarray}
It can be shown that
\begin{eqnarray}\label{qingyuanzi}
\left[R'_x,L_z\right]&=&-iR'_y,\\
\left[R'_y,L_z\right]&=&iR'_x,\nonumber\\
\left[R'_x,R'_y\right]&=&i\left(-2H+2\lambda L_z^2+\frac{1}{4}\lambda-2k\lambda\right) L_z.\nonumber
\end{eqnarray}
These equations imply that ($L_z,R'_x,R'_y$) constitute an polynomial Higgs algebra $SO(3)$ in Hilbert space spanned by degenerate states with a
given energy eigenvalue $E_n$. But the identities in (\ref{rh})
hold only at the aphelion and the perihelion points.
Therefore,  in
general, though the $SO(3)$ dynamical symmetry of two-dimensional spherical hydrogen
atom is broken, it may be restored at
the aphelion and perihelion points of the classical orbits. Because of angular momentum
conservation, the classical orbits are still closed for rational values of $\alpha$.

For the purposes of getting the eigenvalues and the eigenstates simply, we adopt the polar coordinates $(r, \theta)$.
The Hamiltonian of Higgs' systems in a spherical geometry is
\begin{equation}\label{higgs}
\hspace{-1in}H=-\frac{1}{2}\left[3\lambda+\frac{15}{4}\lambda^2r^2+\frac{(1+\lambda
r^2)(1+5\lambda
 r^2)}{r}\frac{\partial}{\partial
 r}+(1+\lambda
 r^2)^2\frac{\partial^2}{\partial
 r^2}+\frac{1}{r^2}\frac{\partial^2}{\partial\theta^2}+\lambda\frac{\partial^2}{\partial\theta^2}\right]+V.
\end{equation}
 For a radial potential $V=V(r)$, the
eigenfunction of energy can be written as
\begin{equation}
\Psi(r,\theta)=e^{im\theta}\psi(r),
\end{equation}
with $m=0, \pm1, \pm2,...$ is the eigenvalue of the conserved angular
momentum $L_z$. The Schr\"{o}dinger equation
$
 H\Psi(r,\theta)=E\Psi(r,\theta)
$
reduces to the radial equation
\begin{equation}\label{re-3}
H_1\psi(r)=E\psi(r).
\end{equation}
Here,
the Hamiltonian $H_1$ can be written as
\begin{eqnarray*}\hspace{-1in}
H_1=-\frac{1}{2}\left[(1+\lambda
r^2)^2\frac{d^2}{dr^2}+\frac{(1+\lambda r^2)(1+5\lambda
r^2)}{r}\frac{d}{dr}-\frac{1+\lambda r^2}{r^2}m'^2+3\lambda
+\frac{15}{4}\lambda^2r^2\right]+\frac{1}{r}+\lambda k,
\end{eqnarray*}
where $m'=\sqrt{m^2-2k}$, and  $\lambda k$ is  a real number independent of $r$.
From the above equation, one could find that the eigenstates and the eigenvalues
can be obtained by substituting $m$ with $m'$ in the Coulomb potential system in the  spherical geometry \cite{higgs1979dynamical}.
The eigenenergy is
\begin{equation}\label{rela}
E_{N,m}=\frac{1}{2}\lambda (m'+N)(m'+N+1)-\frac{1}{2}(m'+N+\frac{1}{2})^{-2}+\lambda k,
\end{equation}
where $N$ is a non-negative integer. But $\Delta m=\pm1$ does not imply  $\Delta m'=\pm1$, the degeneracy of energy splits. And the eigenstates $\Psi(r,\theta)=e^{im\theta}\psi(r)$, where $\psi(r)$ is the same as that in \cite{higgs1979dynamical}, except
 that  $m'$ takes the place of $m$. The above result  (\ref{rela})
reduces to Higgs' Coulomb potential described in \cite{higgs1979dynamical} as $k\rightarrow0$,
 and to the screened Coulomb potential described in \cite{wu2000dynamical} as  $\lambda\rightarrow0$.

\section{Screened isotropic harmonic oscillator}\label{isotropic harmonic}

In this section, we shall discuss the screened spherical isotropic harmonic
 oscillator described by the Hamiltonian (\ref{Hamilton}) with the  potential \cite{wu1998extension}
\begin{eqnarray}\label{harmonic}
V(r)=\frac{1}{2}r^2-\frac{k}{r^2}.
\end{eqnarray}
Similarly, based on the results in \cite{wu2000dynamical} and together with equation (\ref{guiji}),
 the orbit equation
can be expressed as
\begin{eqnarray}
\frac{1}{r^2}=\frac{1}{L_z^2 \alpha^2}\left[E-\frac{1}{2}\lambda L_z^2+\sqrt{(E-\frac{1}{2}\lambda L_z^2)^2-L_z^2 \alpha^2}\cos{(2\alpha(\theta-\theta_0))}\right],
\end{eqnarray}
where $\alpha=\sqrt{1-2k/L_z^2}<1$. As $\lambda$ tends to 0, the equation
reduces to the screened Coulomb potential on a plane. In this situation, the orbit is not closed in general.
But, for rational values of $\alpha$, there still exist an infinite
number of closed orbits.

For the isotropic harmonic oscillator in a spherical geometry, there exist the conserved quantities $L_z,  Q_{xy}=xy+\frac{1}{2}(\pi_x\pi_y+\pi_y\pi_x),$ and  $Q_1=\frac{1}{2}[(x^2-y^2)+(\pi_x^2-\pi_y^2)]$
\cite{higgs1979dynamical}. Combining
the result in \cite{wu2000dynamical}, we introduce the quantities
\begin{eqnarray}
Q_{xy}'&=& (1+\frac{2k}{r^4})xy+\frac{1}{2}(\pi_x\pi_y+\pi_y\pi_x),\\
Q_{1}'&=&\frac{1}{2}\left[(1+\frac{2k}{r^4})(x^2-y^2)+(\pi_x^2-\pi_y^2)\right].\nonumber
\end{eqnarray}
It can be shown that $[Q_{xy}',H]=0$ and $[Q_{1}',H]=0$ hold
at the aphelion (perihelion) points of classical orbits $(\dot{r}=0)$.
The quantities $(L_z,Q_{xy}',Q_1')$
still constitute an polynomial Higgs algebra,
\begin{eqnarray}\label{xiezhenzi}
\left[Q_{xy}',L_z\right]&=&i2Q_1',\\
\left[Q_1',L_z\right]&=&-i2Q_{xy}',\nonumber\\
\left[Q_{xy}',Q_{1}'\right]&=&-i(2+\lambda(2H-\lambda L_z^2))L_z,\nonumber
\end{eqnarray}
which reveal the $SU(2)$ dynamical symmetry as proved in \cite{higgs1979dynamical}.
But the symmetry
holds only at certain points (the aphelion points and the perihelion points) along the classical orbits.

We can get the eigenenergy and the eigenstates
\begin{eqnarray}
E=\frac{1}{2}(1+m'+2N)\left(\sqrt{4+\lambda^2}+\lambda(1+m'+2N)\right)+\lambda k,
\end{eqnarray}
\begin{eqnarray}\label{sjo}
\Psi(r,\theta)=e^{im\theta}r^{m'}(\frac{1}{1+\lambda
r^2})^{\frac{\eta}{\lambda}}F(\alpha,\beta,\gamma,\frac{\lambda
r^2}{1+\lambda r^2}),
\end{eqnarray}
where $m'=\sqrt{m^2-2k}, \eta=\frac{1}{4}(4\lambda+2\lambda
m'+\sqrt{4+\lambda^2}), \alpha=-N, \beta=1+m'+N+\frac{1}{2\lambda}\sqrt{4+\lambda^2},
\gamma=1+m'$, and $N$ is a non-negative integer.
Equation (\ref{sjo}) leads to Higgs' isotropic harmonic oscillator  \cite{higgs1979dynamical}
when $k\rightarrow0$, and to
the screened isotropic harmonic oscillator  \cite{wu2000dynamical}
when $\lambda\rightarrow0$.

\section{Summary}
In this paper, we have shown that, for the two-dimensional spherical screened Coulomb
potential and isotropic harmonic oscillator, there exist
an infinite number of closed orbits for suitable angular momentum
values, and we give the equations of the classical orbits. We construct the extended Runge-Lenz vector for the screened
Coulomb potential and the extended conserved tensor for the
screened isotropic harmonic oscillator. At the aphelion (perihelion) points of classical
orbits, the extended Runge-Lenz vector and the extended conserved tensor are still conserved.
For the screened two-dimensional spherical Coulomb potential and the isotropic harmonic
oscillator, the dynamical symmetries $SO(3)$ and $SU(2)$
are still preserved at certain points of classical
orbits, which behave as the Higgs algebra shown
in (\ref{qingyuanzi}) and (\ref{xiezhenzi}). The eigenenergy and corresponding eigenstates in these systems are also derived.

\section*{Acknowledgments}
 This work is supported by the National Natural Science Foundation of China (Grant Nos. 11105097, 10975075 and 11175089), the
National Basic Research Program of China (Grant No. 2012CB921900), and the National Research Foundation and Ministry of
Education, Singapore (Grant No. WBS: R-710-000-008-271).
\section*{References}
\bibliography{viral}
\vspace{3cm}

\end{document}